\documentclass[conference]{IEEEtran}
\IEEEoverridecommandlockouts
\usepackage{cite}
\usepackage{subcaption}
\usepackage{amsmath,amssymb,amsfonts}
\usepackage{algorithmic}
\usepackage{algorithm}
\usepackage{graphicx}
\usepackage{textcomp}
\usepackage{xcolor}
\usepackage{tikz}
\usepackage{bm}
\usetikzlibrary{arrows.meta, positioning, fit}
\usepackage{booktabs}
\def\BibTeX{{\rm B\kern-.05em{\sc i\kern-.025em b}\kern-.08em
    T\kern-.1667em\lower.7ex\hbox{E}\kern-.125emX}}
\begin{document}

\title{
BASP: Communication-Efficient Batch-Aware Sequence Parallelism for LLM Training
}

\author{
    \IEEEauthorblockN{Bigyan Ghimire, Jon C. Calhoun}
    \IEEEauthorblockA{
        \textit{Holcombe Department of Electrical and Computer Engineering} \\
        \textit{Clemson University}\\
        Clemson, USA 
    }
}

\maketitle

\begin{abstract}
Long-context reasoning for large language models (LLMs) is becoming increasingly important, but training over long sequences remains challenging due to massive memory and communication requirements. 
Sequence parallelism has emerged as an essential technique for addressing bottlenecks in long sequence LLM training. However, we observe that existing sequence parallelism methods are batch-agnostic and apply uniform sequence partitioning across all batch sizes, resulting in inefficient communication. In this paper, we introduce Batch-Aware Sequence Parallelism (BASP), a sequence parallelism approach that leverages batch structure to reduce communication overhead. BASP exploits batch structure by partitioning GPUs into disjoint sequence-parallel groups according to the micro-batch size.  This design reduces the all-to-all communication group size, thereby localizing communication and improving training efficiency. Experimental results on an NVIDIA A100 cluster show that BASP improves end-to-end training time by up to $1.17-1.31 \times$ in Llama and Qwen models compared to standard sequence parallel baselines, while preserving identical model accuracy and memory usage.
\end{abstract}


\section{Introduction}
Transformer-based large language models (LLMs) have gained wide attention as they have demonstrated exceptional capabilities in long context tasks such as long dialogue processing \cite{brown2020language}, large document analysis \cite{chew2023llm}, processing complex codebases \cite{rando2505longcodebench}, and audio, video, and image processing with multi-modal models \cite{zhang2023video}. For instance, when summarizing a book or analyzing a large document, LLMs require long context reasoning over thousands to millions of tokens in a single prompt. While early transformer models operated on a sequence length of 512--2048 tokens \cite{liu2024lost}, current transformer models operate at context windows exceeding 100K tokens \cite{beltagy2020longformer} \cite{ClaudeOpus}. For example, ChatGPT supports 128K tokens \cite{ChatGPT2026}, while Claude models provide a context window of up to 1M tokens \cite{ClaudeOpus}. This has surged the demand for training LLMs in long context windows.

Distributed training strategies such as data parallelism (DP), pipeline parallelism (PP), and tensor parallelism (TP) do not address the long-sequence memory-scaling bottleneck in the transformer training. Data parallelism, which replicates the model across GPUs and partitions the batch, suffers from per-GPU memory pressure \cite{goyal2017accurate}. Pipeline parallelism partitions layers across devices, reducing per-GPU memory pressure but introducing pipeline bubbles, which cause significant under-utilization of GPUs \cite{huang2019gpipe} \cite{arfeen2025pipefill}. Tensor parallelism partitions individual weight matrices across GPUs \cite{shoeybi2019megatron}, resulting in multiple GPUs executing a general matrix multiplication (GEMM). However, it does not parallelize the sequence dimension, which is the main memory bottleneck for long context training \cite{jacobs2023deepspeed} \cite{li2023sequence}.  
 
Recently, for long-context training, sequence parallelism (SP) has been studied as a promising approach \cite{jacobs2023deepspeed} \cite{fang2024usp} \cite{korthikanti2205reducing} \cite{liu2023ring}. SP partitions the sequence dimension across GPUs, distributing both the sequence length and the corresponding activation memory proportionally to the number of participating GPUs. Megatron Sequence Parallelism (Megatron-SP) \cite{korthikanti2205reducing} extends Megatron-LM tensor parallelism by additionally parallelizing the non-tensor parallel layers like layernorm and dropout layers across the sequence dimension \cite{korthikanti2205reducing}. However, the communication volume in Megatron-SP increases linearly with message size, regardless of the number of
GPUs \cite{jacobs2023deepspeed}. SP introduced by DeepSpeed-Ulysses (Ulysses-SP) \cite{jacobs2023deepspeed}, provides a more scalable approach to long context training. Unlike Megatron-SP, which applies sequence partitioning for some layers, Ulysses-SP partitions the sequence dimension for all the layers, including attention and Multi Layer Perceptron (MLP). For attention computation, it employs an all-to-all communication to distribute sequence partitions across the GPUs and computes attention for different heads in parallel. Then it employs
another all-to-all to redistribute the attention output for the sequence partitions. 

However, as the model scale grows,
the training efficiency becomes increasingly constrained by communication rather than computation \cite {chen2024lins}\cite{ge2025bytescale}. Different parallelism strategies exhibit fundamentally different communication patterns. While existing studies \cite{liu2026taco} \cite{ming2024adtopk} \cite{zhang2026helixpipe} have aimed to mitigate communication overhead in traditional parallelism, communication inefficiency in sequence parallelism has been understudied. 

\begin{figure}[ht!]
    \centering
    \includegraphics[width=\linewidth]{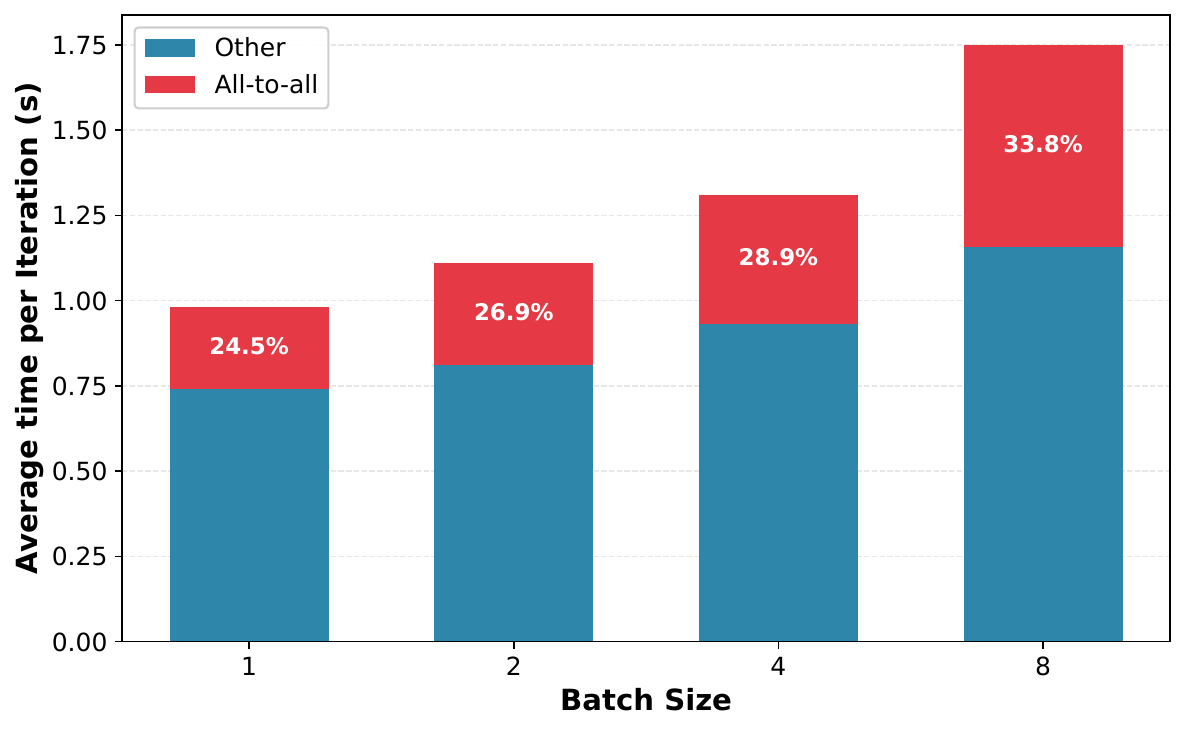}
    \caption{All-to-all's share of time increases linearly 
with micro-batch size (sequence length $= 8K$, $N = 8$, Llama-3.2-1B)}
    \label{fig:alltoall_batch_intro}
\end{figure}

In standard Ulysses, all $N$ GPUs must communicate with $N-1$ peers for all-to-all to redistribute the entire batch\footnote{In this paper, we use batch for convenience, to mean micro-batch, which is the per-GPU fraction of the global batch. } of sequences during attention computation. Our profiling on an 8-GPU NVIDIA A100 
cluster (2 nodes × 4 GPUs/node) reveals that this collective 
becomes a critical bottleneck: as batch size $B$ increases, 
all-to-all time grows proportionally, consuming an increasing 
fraction of total iteration time (Figure~\ref{fig:alltoall_batch_intro}). 

\textbf{Key insights and contributions.} We observe that standard 
Ulysses-SP performs global $N$-way all-to-all regardless of micro-batch 
size $B$ during attention computation. 
However, when $B > 1$ and $N$ is divisible by $B$ 
(i.e., $N = KB$), this 
global communication is unnecessary. We instead assign each of 
the $B$ sequences to a disjoint group of $K = N/B$ GPUs, 
decomposing the single $N$-way collective into $B$ independent 
$K$-way collectives executing in parallel. This reduces the number 
of communication peers per GPU from $N-1$ to $K-1$, with no 
increase in per-GPU memory footprint or sequence shard size.

Crucially, on typical clusters where intra-node bandwidth (NVLink) 
significantly exceeds inter-node bandwidth (InfiniBand), setting 
$K$ equal to GPUs per node confines all-to-all traffic to the 
faster intra-node network.

With these observations, this paper makes the following contributions:
\begin{itemize}

\item Based on our observations, we propose \textit{Batch-Aware Sequence Parallelism} (BASP), where a subgroup of GPUs handles a subset of the batch. This replaces global all-to-all communication with smaller intra-group collectives while keeping the per-GPU memory footprint and sequence shard size unchanged.

\item We evaluate our approach on variants of LLama and Qwen model families. Our results show that Llama 3.1-8B achieves a speedup of $1.21\times$ and Qwen 1.5-1.8B achieves a speedup of $1.32 \times$ compared to Ulysses-SP. Our results also show a reduction of all-to-all time by up to $85\times$ in micro-batch size of 8.
\end{itemize}
 
The remainder of this paper is organized as follows. Section II provides background on transformers and parallelism methods. Section III provides the observation and motivation for our work. Section IV  details the design and implementation of BASP. Section V presents experimental results and analysis. Section VI discusses the limitations, Section VII discusses the related works, and Section VIII concludes.

\section{Background}
\subsection{Transformer Architecture}
The transformer architecture \cite{vaswani2017attention} is the foundation of the large language models. At its core, 
the transformer processes input sequences through stacked layers 
of self-attention and feed-forward networks. The self-attention 
mechanism enables each token to attend to all other tokens in the 
sequence, capturing long-range dependencies that are crucial for 
language understanding.

As model and sequence lengths scale, the quadratic complexity 
$O(N^2)$ of attention computation becomes the primary bottleneck 
\cite{dao2022flashattention}. This has motivated various parallelization 
strategies to distribute attention computation across multiple GPUs. 
To understand these strategies, we first review the multi-head 
attention mechanism that underlies all modern transformers. The multi-head attention is based on the scaled dot-product attention.

\textbf{Scaled Dot-Product Attention}: Given $L$ tokens $X \in \mathbb{R}^{L \times d_{in}}$ and $L'$ tokens $X \in \mathbb{R}^{L' \times d_{in}}$ of dimensions $d_{in}$ dimensions each, the scaled dot-product attention is computed as follows:
\begin{equation*}
\begin{aligned}
\text{Attention}(Q, K, V) &= \text{softmax}\left(\frac{QK^T}{\sqrt{d_k}}\right)V \\
\text{with } Q=XW_Q, \quad K&=X'W_K, \quad V=X'W_V
\end{aligned}
\end{equation*}
where $W_Q \in \mathbb{R}^{d_{in} \times d_{k}}$, $W_K \in \mathbb{R}^{d_{in} \times d_{k}}$ and $W_V \in \mathbb{R}^{d_{in} \times d_{v}}$ are the projection matrices for the query, key and values respectively.
In self-attention, the input tokens are the same, i.e, $X=X'$.

\textbf{Multi-Head Attention}: To capture different aspects of the relationship in the tokens, multi-head attention was introduced \cite{vaswani2017attention}. If $h$ is the number of heads, multi-head attention calculates $h$ different low-dimensional projections of the matrices $\textbf{(Q, K, V)}$, performs scaled-dot product attention for each head, concatenates the outputs, and applies a concatenation to the projection output. 

\begin{equation*}
\begin{aligned}
\text{Multihead}(X, X') &= \text{Concat}(H_{1}, H_{2}, \dots, H_{h}) W_O \\
&\text{with } H_i = \text{Attention}(XW^{i}_Q, X'W^{i}_K, X'W^{i}_V)
\end{aligned}
\end{equation*}

where $W^{i}_Q \in \mathbb{R}^{d_{in} \times d_{k}/h}$, $W^{i}_K \in \mathbb{R}^{d_{in} \times d_{k}/h}$ and $W^{i}_V \in \mathbb{R}^{d_{in} \times d_{v}/h}$ are the $i_{th}$ projection matrices for the query, key and values respectively. $W_O \in \mathbb{R}^{d_{dv} \times d_{out}}$ is the final output projection matrix.

\textbf{Memory bottleneck:} The attention computation in transformer models is a major computational and memory bottleneck due to its quadratic complexity with respect to the sequence length. In particular, the attention score matrix $QK^T$ requires $O(N^2)$ memory, which poses a memory constraint for training.

This challenge is especially amplified in long-context distributed training for large language models, where activation memory grows rapidly with increasing sequence length. Even when model parameters are distributed across multiple GPUs, the intermediate attention activations may exceed the memory capacity of a single device.

Techniques like FlashAttention \cite{dao2022flashattention} reduce memory overhead by avoiding explicit materialization of the full attention matrix by using tiling and online softmax attention, reducing memory complexity from $O(N^2)$ to $O(N)$ for intermediate activations. However, despite this improvement, long-context training still encounters memory and communication bottlenecks at scale \cite{yao2025training}, motivating sequence-parallel approaches such as DeepSpeed-Ulysses.

\subsection{Modes of parallelism}

Training large language models at scale relies on multiple forms of parallelism, each targeting a different computational or memory bottleneck.

\textbf{Data Parallelism (DP):}
In data parallelism, the input batch is partitioned across participating GPUs, with each GPU maintaining a full replica of the model parameters. Each device performs forward and backward computation on its local mini-batch, followed by gradient synchronization across all devices. Due to its replication of model parameters  (gradients, optimizer states, and weights) across all devices, training large models is not feasible because a single device cannot hold the whole model.

\textbf{ZeRO / Fully Sharded Data Parallel:}
ZeRO \cite{rajbhandari2020zero} and Fully Sharded Data Parallel (FSDP)\cite{zhao2023pytorch} improve upon standard data parallelism by partitioning model states across devices instead of fully replicating them on each GPU. In ZeRO, optimizer states (ZeRO-1), gradients (ZeRO-2), and model weights (ZeRO-3) are sharded instead of fully replicated. FSDP follows a similar strategy and is functionally closest to ZeRO-3. This type of partitioning reduces the per-GPU memory usage and allows training models that exceed the memory capacity of a single device.

\textbf{Tensor Parallelism (TP):}
Tensor parallelism, introduced by Megatron-LM \cite{shoeybi2019megatron}, partitions the computation within individual layers across multiple GPUs. Commonly, large matrix multiplications in MLP and attention layers are split along the hidden or output dimensions, allowing each GPU to compute a portion of the GEMM operation. In order to reconstruct the full activation, GPUs must exchange intermediate activations using collective communication operations such as all-reduce, all-gather, or reduce-scatter. For attention computation, Megatron-LM
partitions the sequence along sequence dimensions and applies all-gather and reduce-scatter collective to aggregate QKV
projections. While TP reduces memory and compute pressure per device, these communication operations introduce overhead.

\textbf{Pipeline Parallelism (PP):}
Pipeline parallelism (PP) involves partitioning the layers into subgroups of layers known as stages and assigning the stages to different GPUs. Micro-batches are then pipelined through these stages in the forward and backward passes. PP is effective at alleviating the huge memory pressure on a GPU for very deep transformer models, but it introduces pipeline bubbles and scheduling complexity, which lead to underutilization of GPUs and therefore reduced training efficiency.

\textbf{Sequence Parallelism (SP):}
Megatron-SP \cite{korthikanti2205reducing} adds parallelization to layers that were previously not tensor-parallelized in Megatron-LM. The input is partitioned in the sequence dimension for operations such as layer normalization and dropout, while tensor parallelism is still used for linear layers, including attention and MLP. Communication volume in Megatron-SP’s sequence parallelism increases linearly with the sequence length, irrespective of the number of accelerators.

\textbf{Context Parallelism (CP):}
Context parallelism (CP) was introduced by DeepSpeed-Ulysses \cite{jacobs2023deepspeed} to support extremely long context lengths. Unlike standard sequence parallelism, CP distributes the full sequence across devices for attention computation itself, allowing training on context windows that would otherwise exceed device memory limits. It maintains a constant communication volume regardless of the increase in sequence
lengths and device counts, achieving increased training efficiency. 

\subsection{DeepSpeed Ulysses}
DeepSpeed-Ulysses partitions the individual sequence along the sequence dimension among participating GPUs. For micro batch size $B$, sequence length $S$, and number of GPUs $N$, each GPU holds $\frac{B*S}{N}$ tokens. Then, each GPU calculates the $Q, K, V$ matrices for their portion of the sequence tokens independently for all heads. Then, an all-to-all between $S_p = N$ (degree of parallelism) GPUs is initiated, which distributes the $Q, K, V$ projections across various GPUs that are responsible for processing different heads for the full sequence. This allows the heads to be computed in parallel, independently. Then another all-to-all is initiated to redistribute attention output of the tokens back to the GPUs that held the sequence partitions initially, so other layers like MLP, LayerNorm, etc. continue. An example is presented in Figure 2a.

This design allows attention computation to remain mathematically equivalent to full self-attention while reducing per-device memory usage through sequence partitioning, at the cost of an additional all-to-all communication step.

\begin{figure*}[t!]
    \centering
    
    \begin{subfigure}
    {0.49\textwidth}
        \centering
        \includegraphics[width=\linewidth]
          {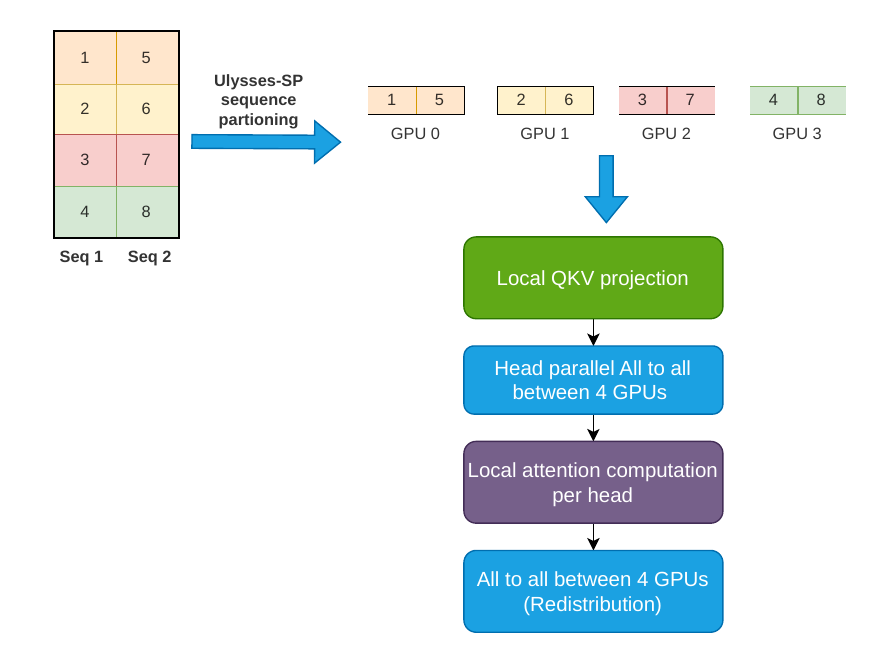}
        \caption{\textbf{Deepspeed Ulysses workflow.}}
    \end{subfigure}
    \hfill
    \begin{subfigure}{0.49\textwidth}
        \centering
        \includegraphics[width=\linewidth]
        {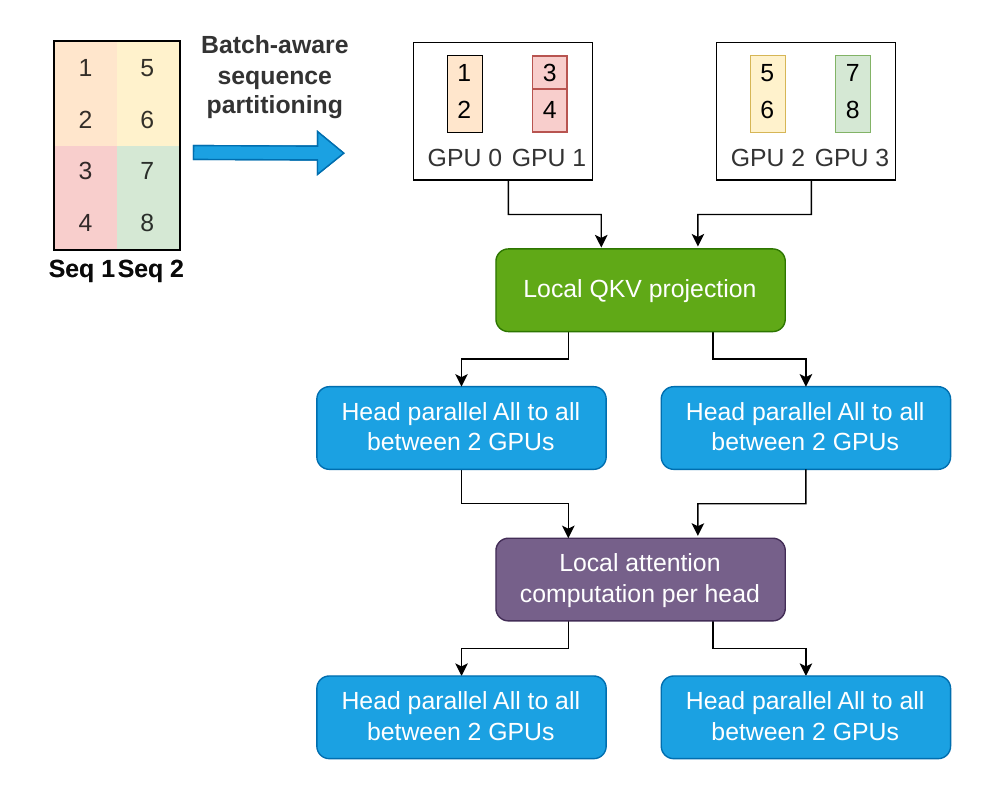}
        \caption{\textbf{Batch-aware Sequence Parallelism 
        (BASP) workflow reducing all-to-all.}}
    \end{subfigure}
  
\caption{\textbf{An example showing the working of Ulysses-SP and BASP in high-level for a micro-batch size of 2 sequences and 4 GPUs}. The sequence is sharded among the GPUs equally. The methods differ in the sequence partitioning pattern. Each GPU then calculates the Q, K, and V matrices for their portion of tokens for all heads. Then, in Ulysses-SP, a single all-to-all happens between 4 GPUs, while in BASP, two all-to-alls happen between 2 GPUs in parallel for attention calculation.  After the attention calculation, the attention output is redistributed back to the original GPUs using another all-to-all. } 
    
    \label{fig:ulysses_vs_ours}
\end{figure*}

\section{Observation and Motivation}

\subsection{Ulysses-SP Communication pattern}
In Ulysses-SP, each sequence in a micro-batch is partitioned among $N$ GPUs ($S_p$ degree), such that each GPU holds a chunk of tokens from all sequences in the micro-batch. During the attention computation, $N$ GPUs participate in the all-to-all for head-parallel attention computation.  This incurs $N$-way all-to-all across GPUs.
 \begin{figure}[ht!]
    \centering
    \includegraphics[width=\linewidth]{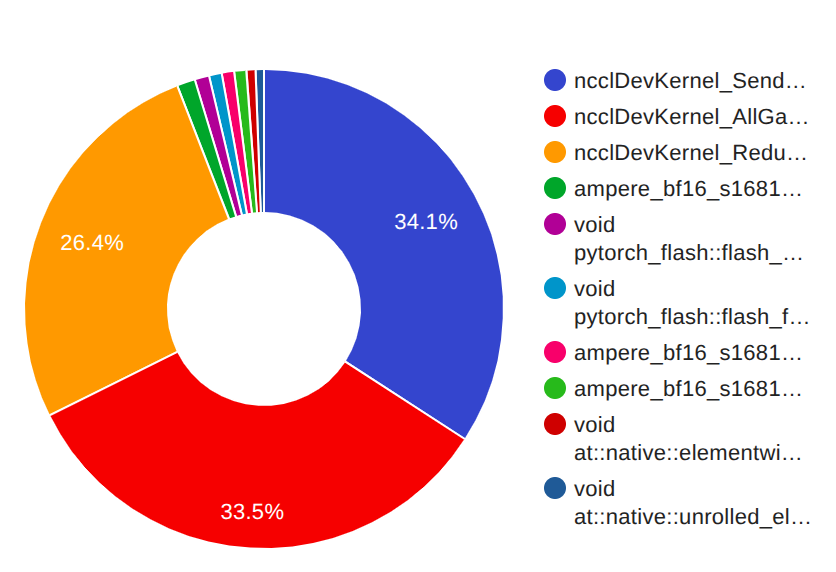}
    \caption{Percentage of time spent during GPU kernel execution on an iteration of SP training (Llama-3.2-3B). \texttt{NcclSend\_Recv(blue)} (used to implement all-to-all in NCCL) shows $34\%$ of total execution time.  }
    \label{fig:alltrend}
\end{figure}

We profile the GPU kernels during a single iteration of sequence parallel training (Figure \ref{fig:alltrend}). Profiling shows that all-to-all accounts for a non-negligible portion of the execution time, i.e., $34\%$. Figure \ref{fig:alltoall_batch_intro}, shows that the all-to-all communication time increases as micro-batch size increases. In a network-constrained cluster, the cost of transmitting these data crossing
the entire cluster cannot be easily hidden via pipelining the communication and computation. 

\textbf{Motivation: }
Our core idea is to exploit the micro-batch size to reduce the number of participants involved in each all-to-all communication, thereby improving training efficiency. This intuition is consistent with classical collective communication models such as the $\alpha$--$\beta$ model, where communication cost consists of a latency term and a bandwidth term.\footnote{Under the standard $\alpha$--$\beta$ communication model, a pairwise-exchange all-to-all can be approximated as 
$T_{\mathrm{a2a}} = (P-1)(\alpha + m\beta)$, 
where $P$ is the number of participating ranks, $m$ is the message size exchanged per round, $\alpha$ is the latency, and $\beta$ is the inverse bandwidth.}
This motivates us to design a new system that
reduces the cost of all-to-all communication while preserving the generality
and usability advantages. By partitioning communication into smaller groups determined by the micro-batch structure, our approach limits synchronization to fewer participants. Our method jointly exploits both the batch and sequence dimensions to reduce the number of communicating devices in a group during attention computation, which allows faster all-to-all communication. This decreases latency overhead and improves communication efficiency, while still preserving the correctness of sequence parallel execution.  We give detailed descriptions of our methodology
in the next section.

\section{Batch Aware Sequence Parallelism}

\textbf{Overview:} BASP exploits micro-batch structure to reduce the all-to-all overhead. Ulysses-SP performs a global $N$-way all-to-all across 
all GPUs to redistribute sequence chunks during attention computation, 
regardless of batch size. When micro-batch size $B$ satisfies $B >1$ and $N = KB$ (where $K$ is an integer), we can 
exploit this structure to reduce communication overhead. Instead of distributing all $B$ sequences among 
all $N$ GPUs, we assign each sequence to a disjoint group of 
$K = N/B$ GPUs. Each group performs an independent $K$-way all-to-all 
on its assigned sequence, reducing per-GPU communication peers from 
$N-1$ to $K-1$ during the attention computation. When $K$ equals the number of GPUs per node, all 
communication remains on fast intra-node interconnects (NVLink), 
avoiding slower inter-node links (InfiniBand) entirely.

Our method spans three modules: (1) Batch-aware group formation, (2) Batch-aware Sequence Partitioning, and (3) Subgroup all-to-all.

\subsection{Batch-aware group formation}
 \begin{figure}[H]
    \centering
    \includegraphics[page=3,width=\linewidth,trim=2cm 2cm 0cm 2cm,clip]{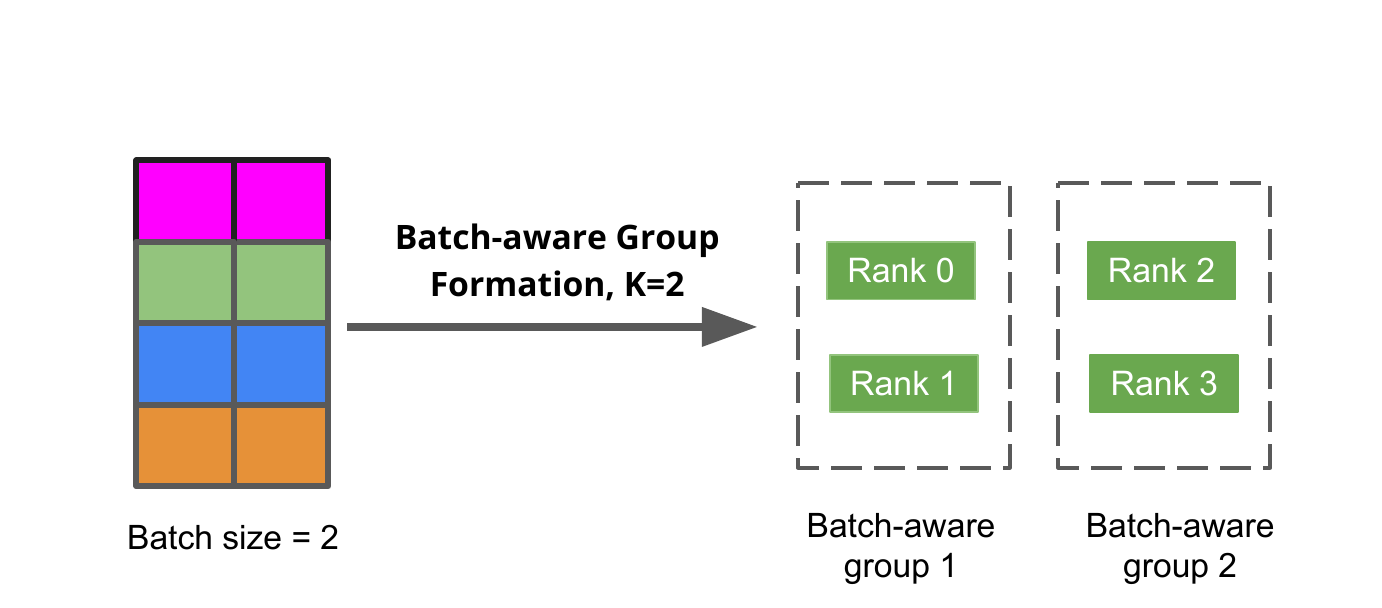}
    \caption{Batch-aware group formation example in $N=4$ GPUs with $B=2$.  }
    \label{fig:batch_aware_group_form}
\end{figure}
We first create multiple batch-aware groups that process the sequences in parallel. While Ulysses-SP creates a single global process group corresponding to the SP degree, BASP partitions the available GPUs into multiple independent subgroups within the SP degree (which we refer to as batch-aware group) at initialization time, with each subgroup responsible for processing a distinct subset of the batch.

Given total world size $N$
 and microbatch size $B$ we compute the group size as:
 \begin{align*}
     K=\frac{N}{B}
 \end{align*}

This formulation ensures that each of the $B$ sequences in the batch is assigned to exactly $K$
GPUs, with no GPU participating in multiple groups (Figure \ref{fig:batch_aware_group_form}). When $B>N$, there are more sequences than GPUs available for partitioning; and thus BASP falls back to standard data parallelism. Currently, BASP assumes that $K=N/B$ is an integer. Extending BASP to support non-divisible configurations is part of our future work. Algorithm \ref{alg:basp_group_creation} shows the complete group formation logic. 
\begin{algorithm}
\caption{Batch-Aware Process Group Creation}
\label{alg:basp_group_creation}
\begin{algorithmic}[1]
\renewcommand{\algorithmicrequire}{\textbf{Input:}}
\renewcommand{\algorithmicensure}{\textbf{Output:}}

\REQUIRE world\_size $N$, batch\_size $B$
\ENSURE Process group for current rank

\STATE $K \leftarrow N / B$ \COMMENT{// Group size}
\STATE $num\_groups \leftarrow B$ \COMMENT{// Number of groups}
\STATE \textbf{assert} $N \bmod B = 0$ \COMMENT{// Validate divisibility}
\STATE 
\STATE $global\_rank \leftarrow \text{dist.get\_rank}()$
\FOR{$i = 0$ \TO $num\_groups - 1$}
    \STATE $ranks \leftarrow [i \cdot K, i \cdot K+1, \dots, i \cdot K+K-1]$
    \STATE $group \leftarrow \text{dist.new\_group}(ranks)$
    \IF{$global\_rank \in ranks$}
        \RETURN $group$ \COMMENT{// This rank's subgroup}
    \ENDIF
\ENDFOR
\end{algorithmic}
\end{algorithm}

\begin{algorithm}
\caption{Batch-Aware Sequence Partitioning}
\label{alg:basp_seq_partition}
\begin{algorithmic}[1]
\renewcommand{\algorithmicrequire}{\textbf{Input:}}
\renewcommand{\algorithmicensure}{\textbf{Output:}}
\REQUIRE Batch $X \in \mathbb{R}^{B \times L \times H}$, group\_size $K$, global\_rank $r$
\ENSURE Local shard $X_{local}$
\STATE $my\_group \leftarrow  \lfloor rank / K \rfloor$
\STATE $my\_local\_rank \leftarrow  rank \bmod K$
\STATE 
\STATE $seq\_chunk\_len \leftarrow L / K$ 
\STATE 
\STATE \COMMENT{// Extract assigned sequence and shard}
\STATE $seq\_start \leftarrow my\_group$
\STATE $seq\_end \leftarrow my\_group + 1$
\STATE $token\_start \leftarrow my\_local\_rank \cdot seq\_chunk\_len$
\STATE $token\_end \leftarrow (my\_local\_rank + 1) \cdot seq\_chunk\_len$
\STATE 
\STATE $X_{local} \leftarrow X[seq\_start:seq\_end, token\_start:token\_end, :]$
\RETURN $X_{local}$

\end{algorithmic}
\end{algorithm}
On multi-node clusters, communication performance depends critically on whether collectives stay within fast intra-node links (NVLink) or must cross slower inter-node network (InfiniBand). To exploit this hierarchy, we must ensure that each group's all-to-all traffic remains entirely within a single node when possible.

We achieve this through contiguous rank assignment: GPUs 0 through $K-1$ form group 0, GPUs $K$ through $2K-1$ form group 1, and so forth. When $K$ equals the number of GPUs per node, this contiguous assignment guarantees each group maps to exactly one physical node, confining all communication to NVLink when possible.

\subsection{Batch Aware Sequence Partioning}
The sequence partioning logic determines how sequences are distributed across GPUs.  In Ulysses-SP, for a microbatch of size $B$ and SP degree $N$, each sequence is partitioned into $N$ contiguous chunks along the sequence dimension, and every GPU processes the corresponding chunk for all sequences in the batch. In contrast, our method jointly exploits both the batch and sequence dimensions. Instead of assigning all GPUs to shard every sequence, we partition the available GPUs into multiple SP groups based on the microbatch size. Then each sequence is only distributed across the GPUs within its assigned group. 

This layout preserves the per-GPU computational load while eliminating the need for cross-sequence aggregation inside the SP group.
As a result, communication becomes localized to the GPUs assigned to each batch element, improving scalability for long-context training.

\textbf{Process group assignment.} To enable each GPU to determine (1) which sequence(s) it is responsible for, and (2) which shard of that sequence it should process, each rank is assigned to group $g = \lfloor rank / K \rfloor$ with local position $\ell = rank \bmod K$, where $K$ is the group size.

Algorithm \ref{alg:basp_seq_partition} shows the complete batch sharding procedure executed during data loading.

\begin{figure}[h]
    \centering

    \begin{subfigure}{\linewidth}
        \centering
        \includegraphics[page=1,width=\linewidth]{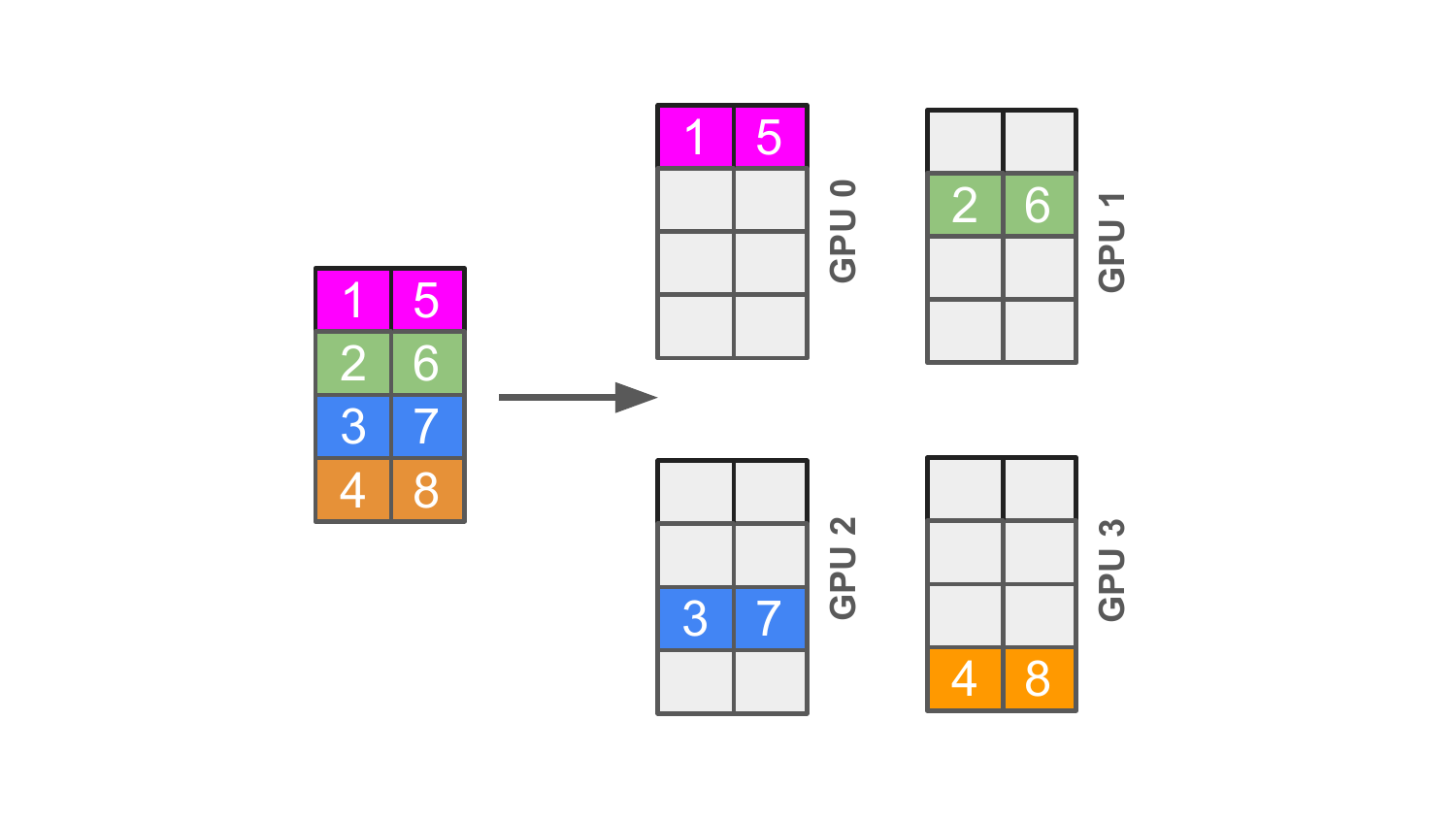}
        \caption{Deepspeed-Ulysses SP sequence partitioning}
    \end{subfigure}

    \vspace{0.5em}

    \begin{subfigure}{\linewidth}
        \centering
        \includegraphics[page=2,width=\linewidth]{Image/Methodology/data_distribution.pdf}
        \caption{BASP (ours) sequence partitioning}
    \end{subfigure}

    \caption{Example showing a) Ulysses-SP and b) BASP (ours) sequence partitioning for $N=4$ GPUs and microbatch size $B=2$, with each sequence of length $S$. With b), we end up assigning a complete sequence to fewer GPUs, which results in less number of participants in all-to-all during self-attention.
}
    \label{fig:alltrend}
\end{figure}

\subsection{Subgroup all-to-all}
\begin{figure}[h]
    \centering

    \begin{subfigure}{\linewidth}
        \centering
        \includegraphics[page=5,width=\linewidth]{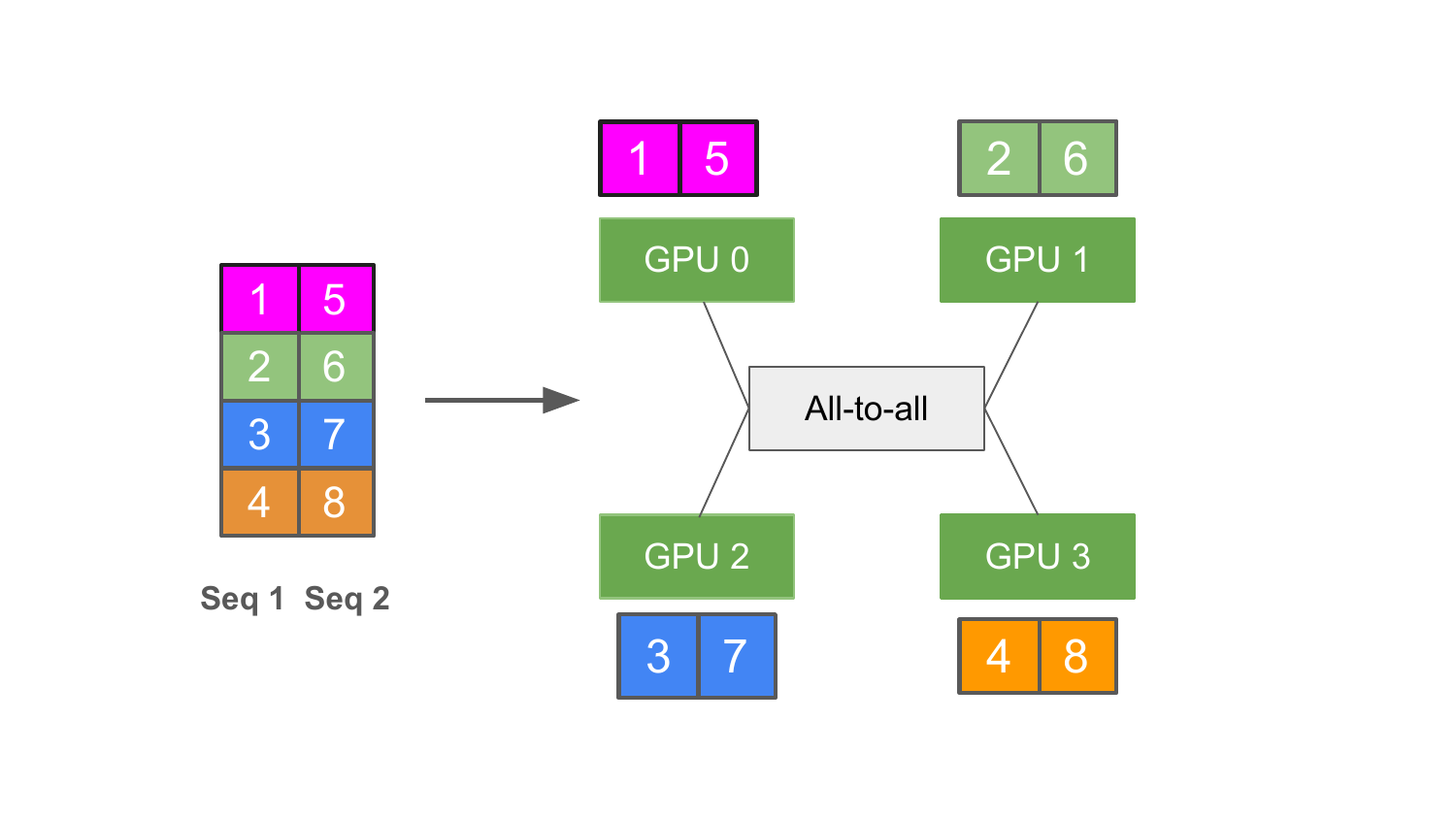}
        \caption{Ulysses-SP all-to-all}
    \end{subfigure}

    \vspace{0.5em}

    \begin{subfigure}{\linewidth}
        \centering
        \includegraphics[page=6,width=\linewidth]{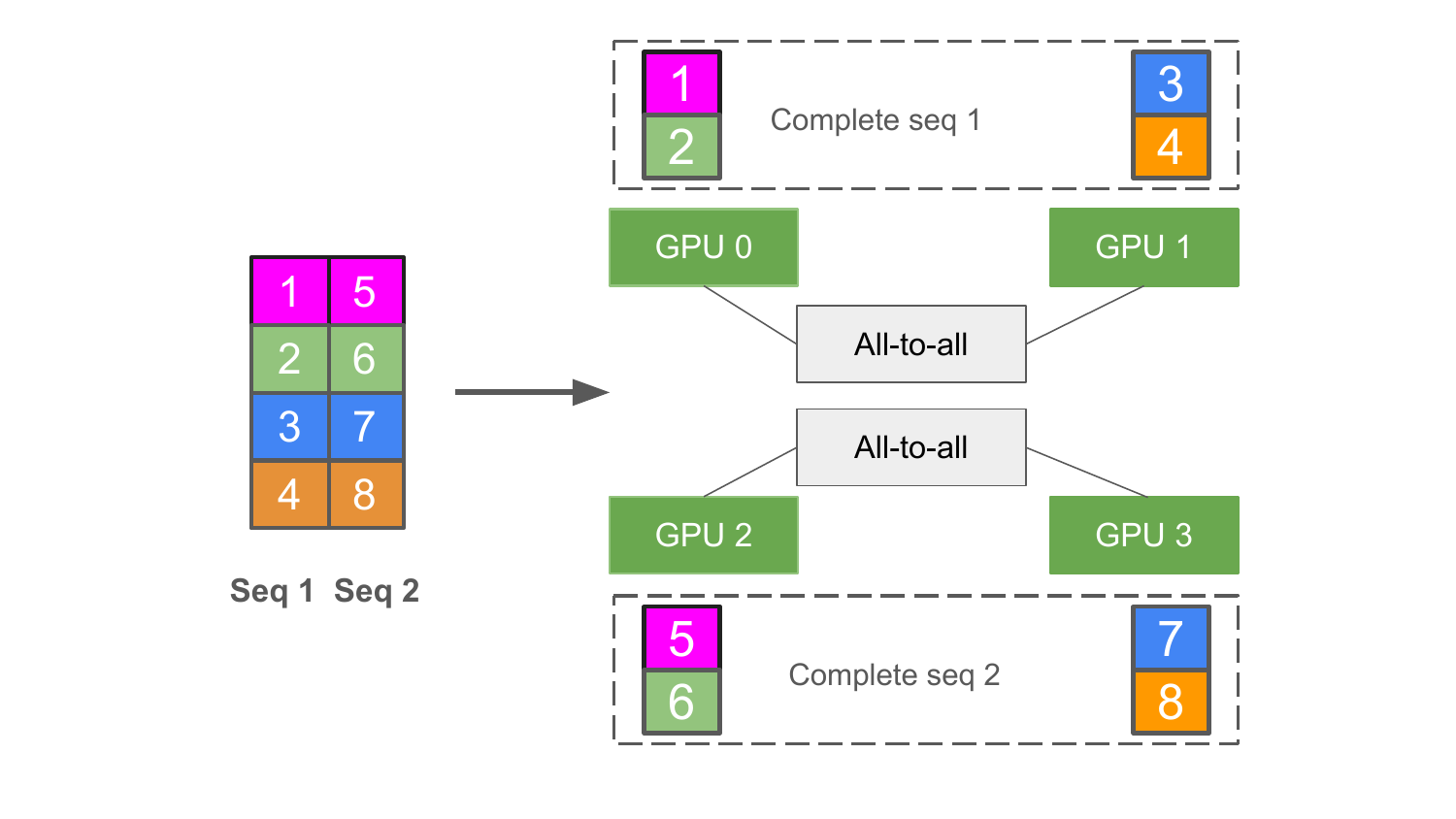}
        \caption{BASP all-to-all}
    \end{subfigure}

    \caption{Example showing a) Ulysses-SP and b) BASP (ours)  all-to-all for $N=4$ GPUs and microbatch size $B=2$, for $seq \text{1}$ and $seq2$. With BASP, 2 GPUs end up holding each complete sequence, resulting in 2 parallel all-to-all between 2 GPUs compared to a single all-to-all between 4 GPUs in Ulysses-SP .
}
    \label{fig:all_all_comparison}
\end{figure}

The attention computation requires all-to-all communication to redistribute $QKV$ projections across attention heads. In Ulysses-SP, this all-to-all spans all $N$ GPUs globally. In BASP, each all-to-all is restricted to the $K$ GPUs within a subgroup.

Our data partitioning, while being computationally similar, reduces the all-to-all communication. With our approach, as shown in Figure \ref{fig:all_all_comparison},  instead of performing a single $N$-way all-to-all across all GPUs, we perform multiple independent $K$-way all-to-alls within GPU groups ($K = N/B$). Each group processes a subset of the batch, with tokens distributed among the $K$ GPUs in that group. 

\subsubsection{Workload and Memory Equivalence.}
In BASP, while a GPU ends up processing greater number attention heads then Ulysses-SP, the computational workload and memory remains same. Ulysses-SP partitions $S$ across $P$ GPUs, giving per-GPU work:
\[
W_{\text{Ulysses}} \propto B \cdot H \cdot \frac{S}{P} \cdot d.
\]

In BASP, the $P$ GPUs are divided into $G$ groups. Each GPU processes $B/G$ batch elements and a sequence shard of size $S/(P/G) = SG/P$, yielding:

\[
W_{\text{BASP}} \propto \frac{B}{G}\cdot H \cdot \frac{SG}{P}\cdot d
= B \cdot H \cdot \frac{S}{P}\cdot d.
\]
Thus, BASP preserves per-GPU workload and the same holds for memory.

\begin{table*}[ht!]
\centering
\caption{Comparison of Parallelism Strategies (for $N$ GPUs, microbatch size $B$ and sequence length $S$)}
\label{tab:parallelism_strategies}
\begin{tabular}{lcc}
\toprule
\textbf{Method} & \textbf{Tokens/GPU} & \textbf{All-to-All Participants} \\
\midrule
Standard Ulysses ($SP=N$) 
& $BS/N$ 
& $N$ GPUs \\

Standard Ulysses ($SP=N/B$) 
& $B^2S/N$ 
& $N/B$ GPUs \\

\textbf{BASP (Ours) $\bm{SP=N, K=N/B}$}
& \textbf{$\bm{BS/N}$}
& \textbf{$\bm{N/B}$ GPUs}  \\

\midrule

\end{tabular}
\end{table*}
\subsection{Combining with ZeRO}
BASP further integrates with ZeRO \cite{rajbhandari2020zero} to further reduce memory consumption in training large language models. ZeRO eliminates memory redundancy by partitioning the model's optimizer states (Zero-1), gradients (Zero-2), and parameters (Zero-3) across data parallel processes.  In BASP, ZeRO partitions model states across both sequence and data parallel groups.

\subsection{Comparison with Sequence Parallel degree }
While the sequence parallelism degree in Ulysses-SP, can be set to $N/B$,  setting $S_p = N/B$ does not replicate BASP. Table \ref{tab:parallelism_strategies} shows the difference in our methods. We see that BASP achieves the benefit of both methods: $S_p = N$ and $S_p=N/B$. The key distinction lies in how sequence length and batch dimensions are distributed across GPUs.

Ulysses-SP with $SP=N$ distributes the sequence length evenly across all $N$ with each GPU processing $BS/N$ tokens. To perform attention, all $N$ GPUs participate in a global all-to-all exchange.

Ulysses-SP provides a \texttt{sequence\_parallel\_size} parameter that allows setting $S_p<N$, partitioning GPUs into groups of size $N/B$. However, this configuration fundamentally changes the workload distribution: each GPU now processes $B^2S/N$ tokens instead of $BS/N$, meaning multiple microbatch sequences are assigned to each GPU rather than sharding individual sequences. While this reduces all-to-all participants to $N/B$ GPUs, it sacrifices the memory efficiency of sequence sharding—sequences are no longer split across GPUs, limiting the maximum trainable sequence length.

BASP achieves the best of both settings: like Ulysses-SP with $S_p=N$, each GPU processes only $BS/N$ tokens, maintaining memory-efficient sequence sharding. However, like the grouped configuration, BASP limits all-to-all participants to $K=N/B$ GPUs by batching sequences within batch-topology-aware groups. This is accomplished by partitioning the $B$ microbatch sequences into $B$
groups of $N/B$
GPUs each, with each group performing independent $K$-way all-to-all operations. The result is reduced communication overhead without compromising per-GPU memory footprint or maximum sequence length capacity.

\section{Experimental Evaluation}
To evaluate the effectiveness of our method, we conduct experiments measuring the end-to-end execution time for representative LLM models, sequence scaling, batch scaling, and all-to-all overhead, and compare to Deepspeed-Ulysses.

\textbf{Evaluation Platform}: We conduct all the experiments on a GPU cluster with 2 nodes, with each node consisting of 4 NVIDIA A100 40GB GPUs connected with NVLink. All nodes are interconnected by 400Gbps InfiniBand network. 

\textbf{Model}: To verify our claims, we perform experiments on variants of two families of models: Llama and Qwen. Table \ref{tab:model_configs} provides the details of the model architectures.
\begin{table}[H]
\centering
\caption{Structure of language models. Llama 3.2B means Llama with 3 billion parameters, and similar for other models.}
\label{tab:model_configs}
\begin{tabular}{lccc}
\toprule
\textbf{Model}  & \textbf{Hidden Size} & \textbf{\# Attention Heads} & \textbf{\# Layers} \\
\midrule
Llama 3.2 1B     & 2048 & 32 & 16 \\
Llama 3.2 3B      & 3072 & 32 & 28 \\
Llama 3.1 8B      & 4096 & 32 & 32 \\
\midrule
Qwen 1.5 1.8B   & 2048 & 32 & 24 \\
Qwen 2.5 3B       & 2560 & 32 & 32 \\
Qwen 3 8B       & 4096 & 32 & 32 \\
\bottomrule
\end{tabular}
\end{table}

\textbf{Implementation and training setup}: We use Deepspeed-Ulysses as our baseline for comparison. We implement our approach by modifying the open-source DeepSpeed code. Our implementation preserves Deepspeed's existing API while introducing batch-aware process group management and data distribution. We use Zero-3 and mixed-precision training for memory savings. The reported metrics are averaged over 30 iterations. 

\subsection{End-to-end execution time}
In this section, we present the end-to-end performance (average per-iteration step-time) comparison of BASP with Ulysses-SP applied to two families of models: Llama and Qwen. We train on a sequence length $16K$ with a batch size of 2 on 8 GPUs. We use micro-batch size of 2 because it is the maximum batch size that allows us to test the performance of our method against Ulysses-SP for sequence length 16K without OOM (Out-of-memory) issues.
\begin{figure*}[t]
    \centering
    
    \begin{subfigure}{0.48\textwidth}
        \centering
        \includegraphics[width=\linewidth]{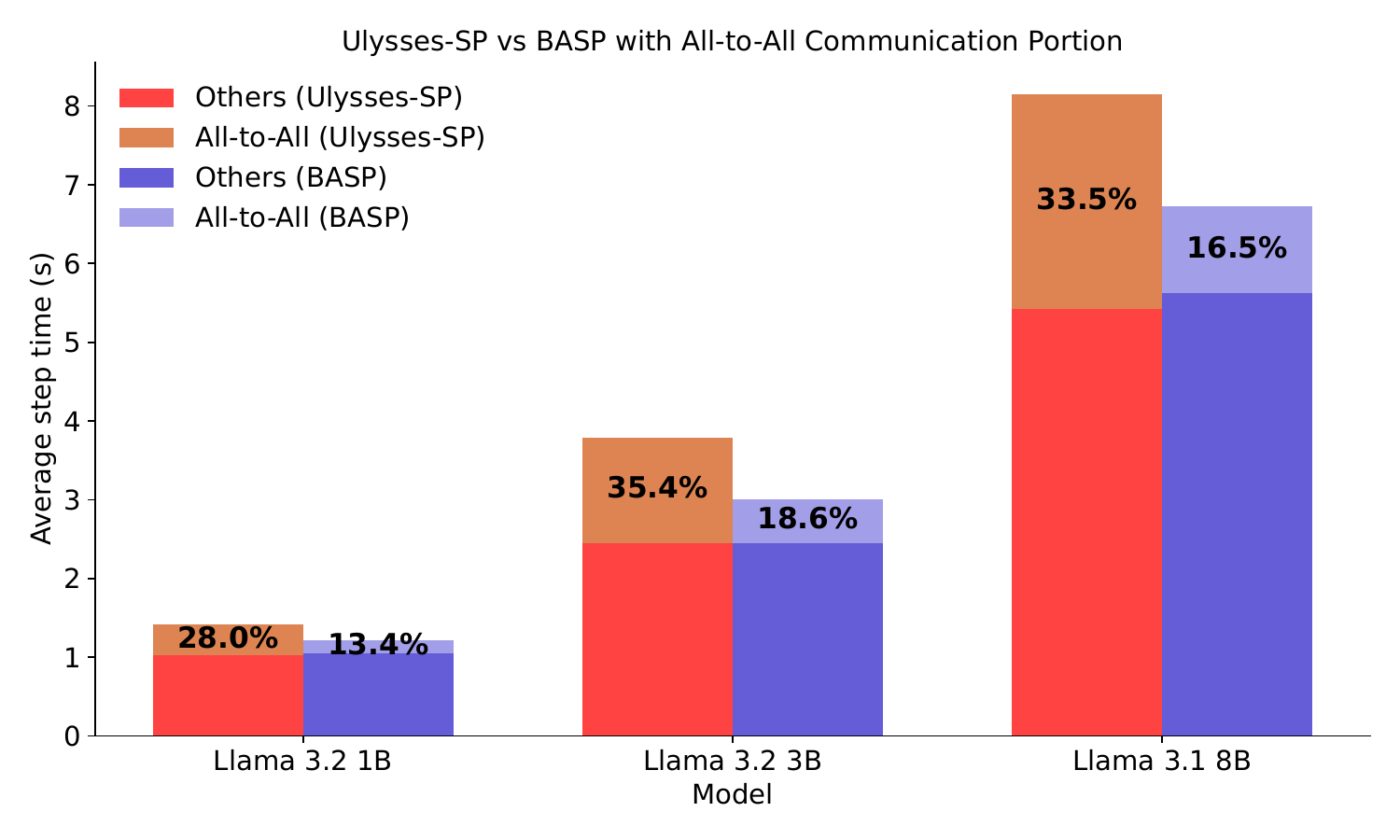}
        \caption{Llama step time}
        \label{fig:llama_throughput}
    \end{subfigure}
    \hfill
    \begin{subfigure}{0.48\textwidth}
        \centering
        \includegraphics[width=\linewidth]{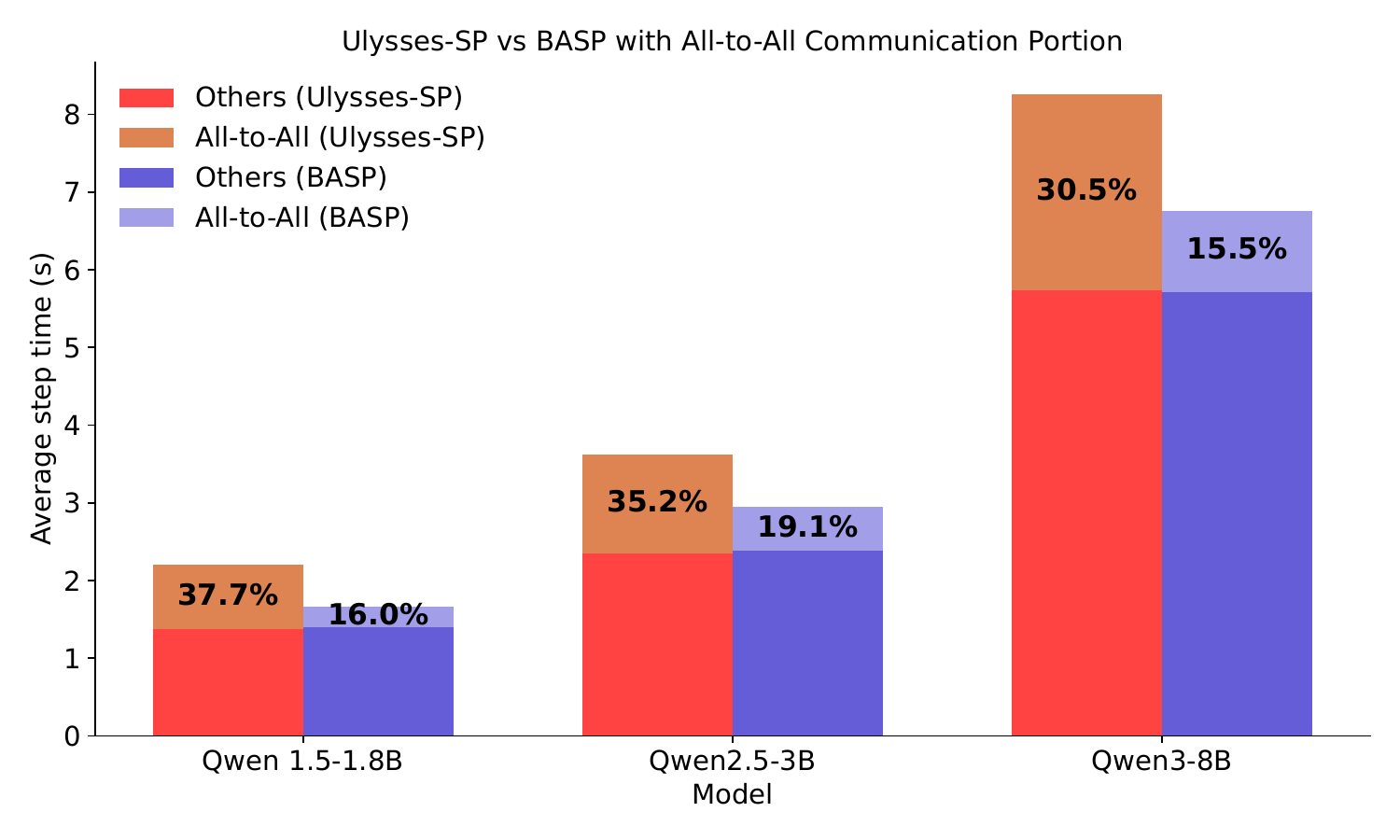}
        \caption{Qwen step time}
        \label{fig:qwen_throughput}
    \end{subfigure}

    \caption{End-to-end step time comparison between Llama and Qwen models}
    \label{fig:end_to_end_result}
\end{figure*}

Figure \ref{fig:end_to_end_result} shows that BASP consistently outperforms Ulysses-SP in all the variants of the model, achieving speedups ranging from $1.17\times$ to $1.32\times$. In the Qwen model, BASP reduces step time by $18.2–24.1\%$ across variants of models, with the largest gain on the 1.8B model: $24\%$ (speedup of $1.31 \times$). In the Llama model, BASP reduces step time by $14.7–20.5\%$ across variants of models, with the largest reduction on the 3.2B model $(20.5\%)$.

\textbf{Communication overhead analysis:}
We measure the aggregate all-to-all communication time per training iteration to isolate the impact of BASP's partitioning strategy. BASP reduces all-to-all communication time by $2.23\times$ to $3.10\times$ across all models. The Qwen 1.5-1.8B model achieves the largest reduction of all-to-all portion from 37.7\% to 16\% ($3.10\times$ all-to-all speedup, $1.31\times$ end-to-end speedup), while Qwen2.5-3B shows the least reduction from 35.2\% to 19\%. ($2.26\times$ all-to-all speedup, $1.22\times$ end-to-end speedup).
These communication reductions exceed the end-to-end speedups, confirming that all-to-all operations constitute only a fraction of the total training time. For instance, in Llama 3.2-1B, despite a $2.45\times$ reduction in communication, the end-to-end speedup is $1.17\times$ because communication accounts for approximately 13.4\% of the total iteration time under Ulysses-SP.
The observed $2.4–3.1\times$ communication reduction aligns closely with our theoretical prediction of approximately $2\times$ for $K=4$ subgroups. The communication benefit is also amplified in BASP because when $K=4$, 4 GPUs are confined to a single node, and during the all-to-all, there is no inter-node communication. 

\subsection{Scaling with Microbatch Size}
\begin{figure}[ht!]
    \centering
    \includegraphics[width=\linewidth]{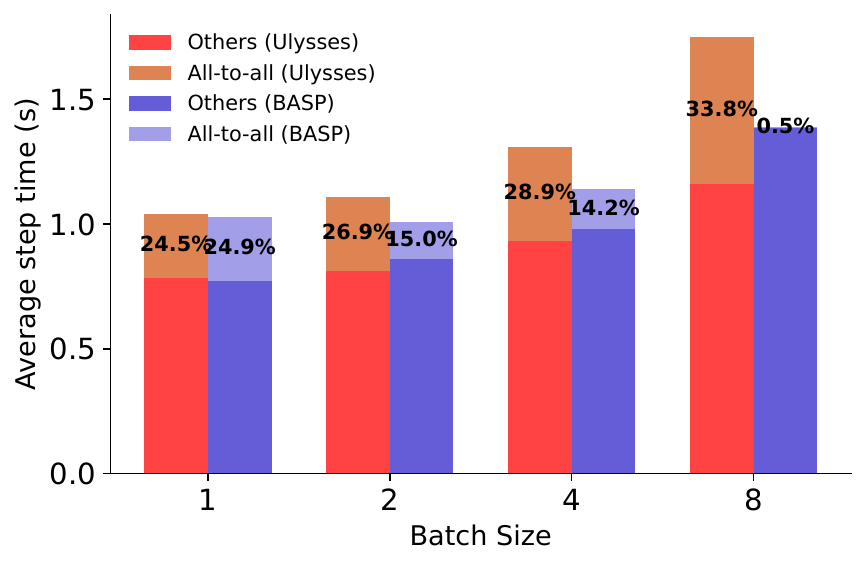}
    \caption{Comparison of batch scaling results for our method (Llama-3.2-1B)}
    \label{fig:batchscaling}
\end{figure}

A core prediction of our method is that speedup should scale with batch size $B$, since larger the $B$, the smaller the group size $K = N/B$. We verify this by fixing sequence length at 8K and varying micro-batch size from 1 to 8 for Llama 3.2-1B. Results are shown in Figure~\ref{fig:batchscaling}.

At batch size 1, BASP and Ulysses-SP are equivalent: $K = N/B = 8/1 = 8$, so both methods perform an identical 8-way all-to-all, hence their step-time is almost identical. At $B \geq 2$, where $N \leq B$, BASP's topology-aware grouping becomes effective, with speedup growing from $1.10\times$ to $1.26\times$ as batch size increases from 2 to 8. These results demonstrate that BASP scales favorably with increasing micro-batch size.

\textbf{Communication overhead analysis}: The performance benefit for BASP comes from the reduced all-to-all communication overhead as seen in Figure \ref{fig:batchscaling}. At B=2, the all-to-all percent increases to 26.9\% for Ulysses-SP while it decreases to 15\% for BASP, and BASP achieves an end-to-end speedup of $1.1\times$. At $B=8$, the all-to-all speedup of BASP is $85\times$ but the end-to-end speedup is $1.25 \times$  compared to the Amdahl's Law theoretical maximum of $1.5\times$
(given Ulysses-SP's 33.8\% all-to-all fraction). This denotes that as batch size increases, and the all-to-all communication becomes negligible (0.5\% at $B=8$), the performance is getting bound by the other collective operations and computation rather than all-to-all communication. Upon profiling BASP, at $B=8$, we find that ZeRO-related communication operations become the bottleneck after sequence parallel all-to-all optimization.

\subsection{End-to-End Performance for Sequence Scaling}
\begin{figure}[ht!]
    \centering
    \includegraphics[width=\linewidth]{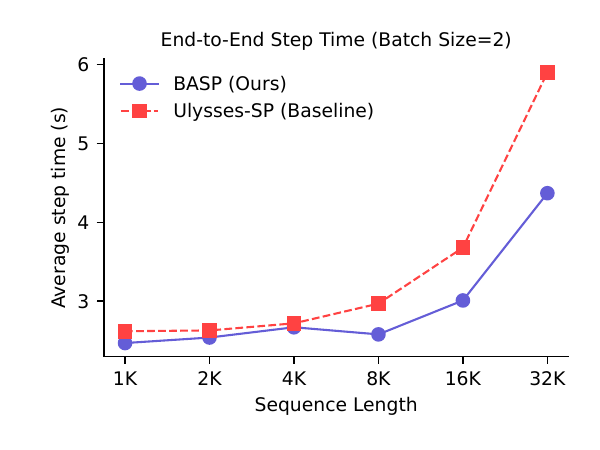}
    \caption{Sequence scaling comparison (Llama 3.2-3B).}
    \label{fig:end_end_seq_speedup}
\end{figure}

In the previous experiments, we showed that our method consistently performs better on the family of models. In this experiment, we show that our method scales with longer sequences. We evaluate the end-to-end time of BASP with Ulysses-SP on sequence lengths up to 32K with a batch size of 2 Llama 3.2-3B. To provide a fair comparison, we use the same setting for both methods. Figure \ref{fig:end_end_seq_speedup} shows that BASP consistently outperforms Ulysses-SP for the sequence
length that can be run with both. 

Across all settings, BASP consistently outperforms Ulysses-SP, with benefits becoming more pronounced at longer sequence lengths. At short sequences (1K–4K), the improvements are modest, with BASP achieving 5.7\% and 3.4\% reduction in step time at 1K and 2K, respectively, and a marginal 1.8\% improvement at 4K. This is expected, as communication overhead is relatively small compared to computation in smaller sequence lengths.
However, as sequence length increases, BASP shows significantly stronger gains. At 8K tokens, BASP reduces step time by 13.1\%, which further increases to 18.2\% at 16K and reaches 25.9\% at 32K. 

These results indicate that BASP provides limited but consistent gains at small sequence lengths while delivering substantial improvements in long-context regimes, where all-to-all communication dominates end-to-end training time.

\subsection{Loss Comparison Plot}
\begin{figure}[ht!]
    \centering
    \includegraphics[width=\linewidth]{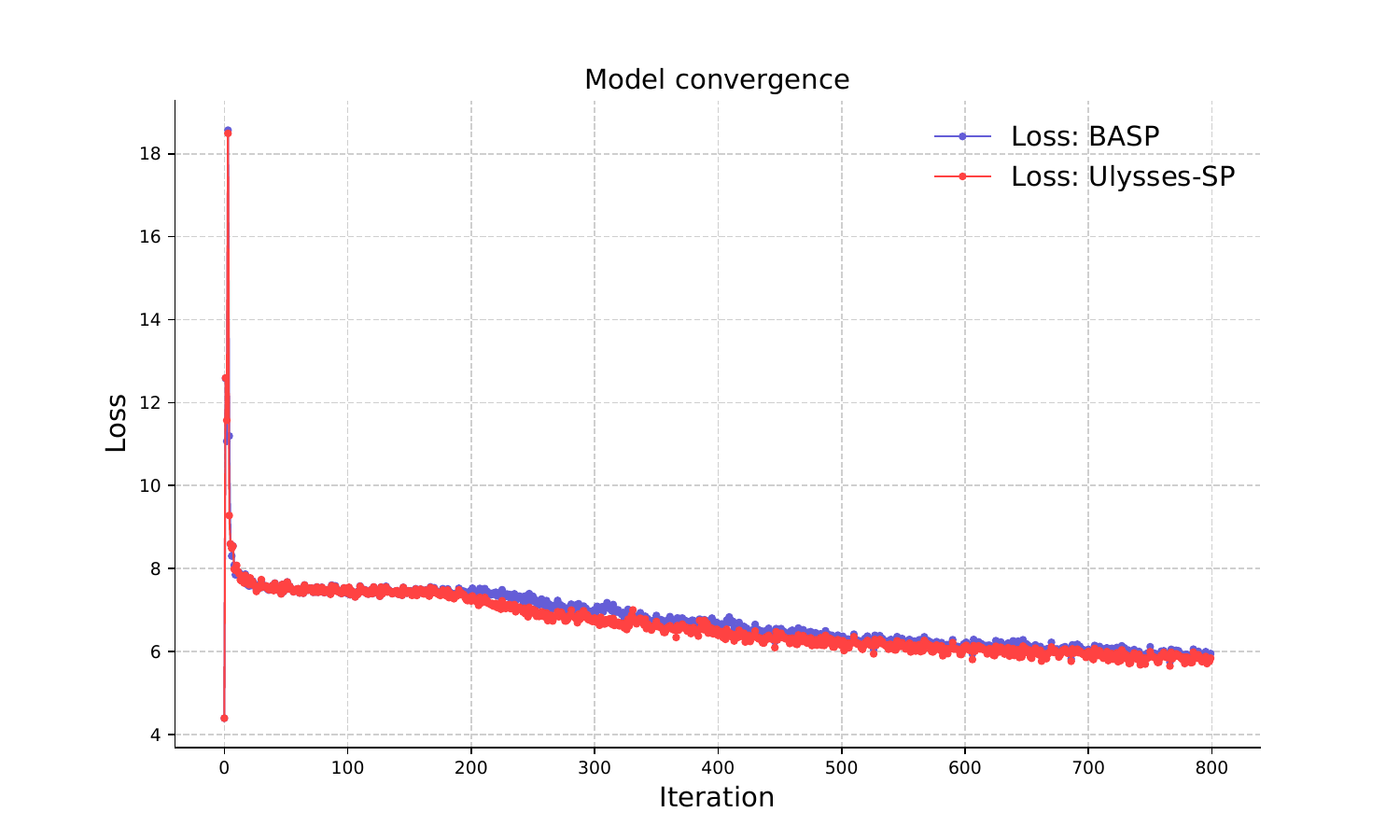}
    \caption{Loss comparison plot showing almost complete overlap for Ulysses-SP and BASP on Llama 3.2-3B, verifying our implementation.}
    \label{fig:loss_plot}
\end{figure}

To validate that our method is purely a communication optimization technique with no impact on accuracy, we compare the loss convergence of BASP and Ulysses-SP. Using the same dataset, we train Llama 3.2-3B for 800 iterations on 8 GPUs. Figure \ref{fig:loss_plot} shows that the loss curves for BASP and Ulysses-SP completely
overlap, which validates that our method preserves the accuracy.

\section{Limitations}
BASP is designed for scenarios where the 
number of GPUs $N$ exceeds or equals the micro-batch size $B$ and $N=KB$ and $K$ is an integer. This 
condition holds for typical long-context training workloads on 
multi-node clusters, where sequence length (and hence memory pressure) 
necessitates distributing each sequence across multiple GPUs. Extending BASP to support non-divisible configurations and more flexible group formation strategies is part of our future work.

\section{Related Works}
There has been a surge of research on long context training in recent years as the scaling of the transformer architecture has been bottlenecked by the memory constraints. Megatron Sequence Parallelism (SP) \cite{korthikanti2205reducing} extends Megatron-LM\cite{shoeybi2019megatron}, a tensor-parallel training framework, by partitioning the dropout and layernorm layers, allowing them to be distributed across multiple GPUs and therefore further reducing the memory pressure of any single GPU. Unlike Megatron-LM, which partitions attention heads across GPUs while replicating the full sequence activations on each GPU, Megatron SP additionally parallelizes along the sequence dimension and uses all-gather before attention and reduce-scatter afterward to reconstruct and redistribute the sequence partitions.

Deepspeed Ulysses\cite{jacobs2023deepspeed} uses all-to-all collective among the GPUs to gather the QKV projections and performs a per-head attention computation. Unlike Megatron-SP, communication analysis shows that DeepSpeed-Ulysses keeps communication volume consistent by increasing GPUs proportional to the sequence length. However, it fails to address the all-to-all communication bottleneck for larger batch sizes.

Ring Attention \cite{liu2023ring} computes attention in a blockwise manner by leveraging the online softmax method to compute the exact attention incrementally. This approach enables training on sequences far exceeding single-device memory capacity. However, Ring Attention faces bottlenecks from peer-to-peer (P2P) communication required for transferring KV chunks, particularly when computation can no longer overlap with communication on long sequences. Striped Attention\cite{brandon2023striped} improves upon Ring Attention by interleaving computation and communication more effectively, achieving better pipeline utilization through careful scheduling of block transfers.

Unified Sequence Parallelism (USP) \cite{fang2024usp} combines DeepSpeed-Ulysses and Ring Attention to counter each method's individual inefficiencies by using all-to-all collectives for short-to-medium sequences where communication can be effectively overlapped, and switching to ring-based communication for extremely long sequences where blockwise computation becomes necessary. However, Unified SP does not optimize the all-to-all collective for larger batch sizes. FlexSP\cite{wang2025flexsp} introduces adaptive switching between different SP strategies based on runtime profiling, dynamically selecting between Ulysses-style all-to-all, ring-based, and hybrid modes depending on sequence length, batch size, and network conditions. This adaptivity comes at the cost of runtime overhead for strategy selection and potential load imbalance during transitions.

MiCS (Mixed-Precision Communication Scheduler) \cite{zhang2022mics} introduces heterogeneous sharding granularities for optimizer states, gradients, and parameters in ZeRO \cite{rajbhandari2020zero}, reducing communication volume by selectively applying mixed precision to different component types. While MiCS optimizes memory and communication for data parallelism, it does not address sequence parallelism or the topology-aware scheduling of all-to-all collectives.

\section{Conclusion}
We presented \textit{Batch-Aware Sequence Parallelism (BASP)}, a simple yet effective optimization to Ulysses-SP that leverages batch structure to optimize global all-to-all communication. By decomposing a global $N$-way collective into independent $K$-way subgroup collectives, BASP reduces the number of communication phases from $(N-1)$ to $(K-1)$ while preserving the per-GPU memory footprint and sequence partition size. Experiments on Llama and Qwen models show that BASP achieves a speedup of $1.21\times$ on Llama 3.1-8B and $1.32 \times$ on Qwen 1.5-1.8B compared to Ulysses-SP with a reduction of all-to-all time by up to 67.7\% in Qwen 1.5-1.8B.





\section*{Acknowledgment}

This material is based upon work supported by the United States Department of Energy under grant DE-SC0026344 and National Science Foundation under Grant No. SHF-1943114. 
Clemson University is acknowledged for generous allotment of compute time on the Palmetto cluster.

\bibliographystyle{plain}
\bibliography{bibliography.bib}

\end{document}